\documentclass[12pt]{iopart}

\usepackage{iopams,setstack}
\usepackage{epsf}
\usepackage{cite}

\newcommand{\bd}{\begin{displaymath}}
\newcommand{\ed}{\end{displaymath}}
\newcommand{\be}{\begin{equation}}
\newcommand{\ee}{\end{equation}}
\newcommand{\ba}{\begin{eqnarray}}
\newcommand{\ea}{\end{eqnarray}}

\begin{document}

\paper[Phonon lineshapes in atom-surface scattering]
{Phonon lineshapes in atom-surface scattering}

\author{R. Mart\'{\i}nez-Casado}
\address{Department of Chemistry, Imperial College London,\\
         South Kensington, London SW7 2AZ, United Kingdom}

\author{A S Sanz and S Miret-Art\'es}
\address{Instituto de F\'{\i}sica Fundamental\\
Consejo Superior de Investigaciones Cient\'{\i}ficas\\
Serrano 123, 28006 Madrid, Spain}

\begin{abstract}
Phonon lineshapes in atom-surface scattering are obtained from a simple
stochastic model based on the so-called Caldeira-Leggett Hamiltonian.
In this single-bath model, the excited phonon resulting from a creation
or annihilation event is coupled to a thermal bath consisting of an
infinite number of harmonic oscillators, namely the bath phonons.
The diagonalization of the corresponding Hamiltonian leads to a
renormalization of the phonon frequencies in terms of the phonon
friction or damping coefficient.
Moreover, when there are adsorbates on the surface, this single-bath
model can be extended to a two-bath model accounting for the effect
induced by the adsorbates on the phonon lineshapes as well as their
corresponding lineshapes.
\end{abstract}

\submitto{\JPCM}

\maketitle


\section{Introduction}
\label{sec1}

Bulk and surface phonon dynamics is a very active field of research.
Usually the information about this dynamics arises from scattering
experiments, where the typical observable quantity is the lineshape
associated with a surface excitation, i.e.\ a creation or an
annihilation event.
At present, the available experimental techniques, such as the
surface scattering with He-atom beams \cite{has} or the He spin echo
\cite{hse}, allow to explore higher and higher angular and energy
resolutions.
Therefore, it is important and necessary to have at our disposal
appropriate theoretical formalisms which allow us to extract reliable
information from the experiment, such as phonon lifetimes and frequency
shifts, phonon-dispersion relations or multiphonon backgrounds,
for example.
In this regard, a qualitative explanation was formerly provided for
multiphonon scattering processes in terms of a semiclassical formula
for energy and momentum exchange within the so-called trajectory
approximation as a generalization of the Brako-Newns formula
\cite{brako,himes}.
Levi and Bortolani \cite{levi} also developed a general multiphonon
formalism based on the time-evolution approach.
More recently, Gumhalter \cite{branko} has made a revision on single
and multiphonon atom-surface scattering within the quantum regime.
However, as far as we know, little attention has been paid to the
theory of phonon lineshapes.

In this work we develop a theory of phonon lineshapes within a
stochastic approach based on the so-called Caldeira-Legget Hamiltonian
\cite{caldeira}, which constitutes the paradigm of stochastic dynamics
and system-plus-reservoir approaches.
In particular, we focus on an excited phonon in the presence of a
phonon field and study its relaxation.
To this end, in next Section we shortly review the theory
of Manson and Celli, based on the transition matrix formalism
\cite{celli1,manson,celli2}, in order to introduce the notation and
their main theoretical findings on single-\ and multiphonon scattering.
In Section~\ref{sec3}, the Caldeira-Leggett Hamiltonian is introduced
to develop a theory for damping phonons within a single-bath model
(phonon bath).
This Hamiltonian, which displays a high degree of versatility, has been
used by our group to describe surface diffusion of single adsorbates
\cite{salva1a,salva1b,salva1c} and interacting adsorbates within a
two-bath model (phonon and adsorbate baths) \cite{salva2a,salva2b}.
This Hamiltonian has been applied, also recently, to classical
atom-surface scattering to describe angular distributions
\cite{eli1,eli2} and energy loss spectra \cite{eli3,eli4}.
Finally, in Section~\ref{sec4}, different aspects of this new approach
are discussed and applied to He scattering off a Cu(001) surface
with a 3\% CO coverage at low (surface) temperatures \cite{exp}.


\section{A brief account on the transition matrix formalism}
\label{sec2}

The observable in atom-surface inelastic scattering experiments is
usually the so-called differential reflection coefficient, which gives
the fraction of probe particles (e.g.\ He atoms) scattered off into a
final solid angle $d \Omega_f$ and an energy interval $d E_f$.
At a theoretical level, this observable is obtained as a transition
rate, dividing the incident flux crossing a plane parallel to the
surface and then multiplying by the density of available final states
of the probe particle \cite{celli1,manson,celli2}.
This allows us to express the transition rate as \cite{celli1,manson}
\begin{equation}
\label{eq:rate}
 \fl
 \mathcal{R}({\bf k}_f, {\bf k}_i) =
  \frac{1}{\hbar^2} \int_{- \infty}^{+ \infty}
   e^{- i (\epsilon_i - \epsilon_f) t / \hbar} \ \!
    |\tau({\bf k}_f, {\bf k}_i)|^2
  \sum_{j,l} \langle  e^{- i {\bf k} \cdot [{\bf R}_l + {\bf u}_l(0)]}
    \ \! e^{ i {\bf k} \cdot [{\bf R}_j + {\bf u}_j(t)]} \rangle \ \!
 dt ,
\end{equation}
where $\epsilon_i$ and $\epsilon_f$ are, respectively, the initial
and final particle energies, $\hbar \omega = \epsilon_f - \epsilon_i$
is the energy exchanged in the scattering process, ${\bf k} = {\bf k}_f
- {\bf k}_i$ is the total wave vector exchanged by the scattering
particles, and $\tau_{fi} \equiv \tau({\bf k}_f, {\bf k}_i)$ is a
pairwise transition operator.
The (time-independent) equilibrium position of the $j$th surface unit
cell is given by ${\bf R}_j$, with $j$ being a discrete two-dimensional
variable for each layer.
For simplicity, we will assume that there is only one atom per unit
cell and, therefore, ${\bf u}_j$ corresponds to the displacement of
the $j$th atom with respect to ${\bf R}_j$ due to its vibrational
motion.
In (\ref{eq:rate}), the sums extend over all ${\bf R}_j$ and the
average is over all initial lattice states (once the sum over all
possible final lattice states has been carried out).

Equation (\ref{eq:rate}) has the general form of the fundamental
equation for a large class of scattering treatments involving a
many-body target,
\begin{equation}
\label{eq:rates}
 \mathcal{R}({\bf k}_f, {\bf k}_i) = \frac{2 \pi}{\hbar} \ \!
  |\tau({\bf k}_f, {\bf k}_i)|^2 \ \! S({\bf k}, \omega) .
\end{equation}
This equation is the product of a {\it form factor}, $|\tau({\bf k}_f,
{\bf k}_i)|^2$, for the scattering centers and a {\it dynamical
structure factor}, $S({\bf k},\omega)$, which depends on the average
over scattering center positions.
The latter factor describes, precisely, the lineshapes in the kind of
experiments we are interested in here, and can be expressed as
\begin{equation}
\label{eq:skw}
 S({\bf k},\omega) = \int I({\bf k},t) \ \! e^{i \omega t} dt ,
\end{equation}
i.e.\ as a time Fourier transform of the so-called {\it intermediate
scattering function},
\begin{equation}
\label{eq:skt}
 I({\bf k},t) \equiv \sum_{j,l} e^{- i {\bf k} \cdot {\bf R}_l}
   e^{i{\bf k} \cdot {\bf R}_j}
   \langle e^{ - i {\bf k} \cdot {\bf u}_l(0)}
   e^{ i {\bf k} \cdot {\bf u}_j(t)} \rangle  .
\end{equation}
If the harmonic approximation is valid for any vibrational lattice
mode, then the (vibrational) average in (\ref{eq:skt}) can be easily
carried out by standard methods, which yields
\begin{equation}
\label{eq:harm}
 \langle e^{ - i {\bf k} \cdot {\bf u}_l(0)}
   e^{ i {\bf k} \cdot {\bf u}_j(t)} \rangle
  =  e^{- \Upsilon_{lj} ({\bf k},t)} ,
\end{equation}
where
\begin{equation}
\label{eq:uplj}
 \Upsilon_{lj} ({\bf k},t) = 2 W({\bf k}) -
 \langle [{\bf k} \cdot {\bf u}_l(0)]
         [{\bf k} \cdot {\bf u}_j(t)] \rangle ,
\end{equation}
and
\begin{equation}
\label{eq:DW}
 W({\bf k})
  = \frac{1}{2} \langle [{\bf k} \cdot {\bf u}_j(t)]^2 \rangle
  = \frac{1}{2} \langle [{\bf k} \cdot {\bf u}_j(0)]^2 \rangle .
\end{equation}
The factor $W({\bf k})$ is the argument of the so-called
{\it Debye-Waller} (DW) {\it factor}, $e^{- 2 W({\bf k})}$, which
governs the decay of the scattered intensity as a function of the
surface temperature through the canonical ensemble average.
This contribution to the thermal attenuation of the scattered
intensities arises mainly from the momentum transfer perpendicular
to the surface.

The second term in (\ref{eq:uplj}) accounts for the energy transfer
between the surface and the scattered particles, which is governed
by the displacement autocorrelation function
\begin{equation}
\label{eq:ujl}
 U_{jl} ({\bf k},t) \equiv \langle [{\bf k} \cdot {\bf u}_l(0)]
  [{\bf k} \cdot {\bf u}_j(t)] \rangle .
\end{equation}
Here we are dealing with the problem of inelastic scattering and,
therefore, our interest relies on the evaluation of this displacement
correlation function.
Hence we express the vibrational amplitude of the $j$th scattering
center in terms of surface normal modes as
\begin{equation}
\label{eq:uj}
 {\bf u}_j (t) = \sum_{{\bf Q}, \nu} \lambda_{{\bf Q}, \nu}
  {\bf e}({\bf Q}, \nu)
   \left[ a_{{\bf Q}, \nu} (t) e^{- i {\bf Q} \cdot {\bf R}_j}
  + a_{-{\bf Q}, \nu}^{\dag} (t) e^{i {\bf Q} \cdot {\bf R}_j} \right] ,
\end{equation}
where
\begin{equation}
 \label{eq:at}
  a_{{\bf Q}, \nu} (t) = a_{{\bf Q}, \nu} (0)
   e^{- i \omega_{\nu} ({\bf Q}) t},
\end{equation}
and
\begin{equation}
\label{eq:lambda}
 \lambda_{{\bf Q}, \nu} =
  \sqrt{ \frac{\hbar}{2 N M \omega_{\nu}({\bf Q})}} .
\end{equation}
In these expressions, ${\bf Q}$ is the parallel-phonon momentum, $\nu$
represents additional phonon quantum numbers (e.g.\ the Rayleigh-mode
branch or normal momentum of bulk modes), $M$ is the mass of the
surface atom in the unit cell, ${\bf e}({\bf Q}, \nu)$ is the
polarization vector, and $a_{{\bf Q}, \nu}(t)$ and $a_{-{\bf Q},
\nu}^{\dag}(t)$ are the time-dependent phonon annihilation and
creation operators, respectively.
The polarization vector is related to the frequency distribution
function, $\rho_{\alpha \beta}$, as \cite{celli1,manson}
\begin{equation}
\label{eq:rho}
 \rho_{\alpha \beta} =
  \sum_{{\bf Q}, \nu} {\bf e}_{\alpha}({\bf Q}, \nu)
   \cdot {\bf e}_{\beta} ({\bf Q}, \nu)
   \ \! \delta (\omega - \omega_{\nu} ({\bf Q})) ,
\end{equation}
where $\alpha$ and $\beta$ are the Cartesian components of the
polarization vectors.
From (\ref{eq:uj}) and (\ref{eq:at}), and after performing the
corresponding thermal averages, (\ref{eq:ujl}) becomes
\begin{eqnarray}
\label{eq:ujlT}
 U_{jl}({\bf k},t) & = & \sum_{{\bf Q},\nu} \lambda_{{\bf Q},\nu}^2
  [{\bf k} \cdot {\bf e}({\bf Q},\nu)]^2 \Big\{ n_{\nu}({\bf Q})
  e^{i {\bf Q} \cdot ({\bf R}_l - {\bf R}_j)}
   e^{- i \omega_{\nu}({\bf Q}) t}
 \nonumber \\ & &
 + [n_{\nu} ({\bf Q}) + 1] e^{- i {\bf Q} \cdot ({\bf R}_l - {\bf R}_j)}
  e^{i \omega_{\nu}({\bf Q}) t} \Big\} ,
\end{eqnarray}
where $n_{\nu}({\bf Q}) = [ e^{\hbar \omega_{\nu}({\bf Q}) /
k_B T} - 1]^{-1}$ is the {\it Bose-Einstein factor}.
By Taylor expanding the exponential of $U_{lj}({\bf k},t)$,
with $\Upsilon_{lj}$
decomposed as in (\ref{eq:uplj}), the multiphonon
contributions associated with the total inelastic scattering process
can be obtained.
The first term of this series expansion (namely the unity) gives the
diffraction intensities, i.e.\ the elastic contribution to the
scattering process, after carrying out the integral
(\ref{eq:skw}) and then computing the transition rate (\ref{eq:rates}).
At this level of approximation, as reported by Manson \cite{manson},
these diffraction intensities will be given by the product of three
factors: the DW factor, the form factor and the structure factor.
In particular, the (static) structure factor is the time-independent
factor arising from (\ref{eq:skt}) at zero temperature,
\begin{equation}
\label{eq:sk}
 S({\bf k}) \equiv \sum_{j,l} e^{- i {\bf k} \cdot {\bf R}_j}
  e^{i {\bf k} \cdot {\bf R}_l} ,
\end{equation}
with the diffraction peaks then expressed in term of
$\delta$-functions.
The next term, the linear one, provides the single-phonon contribution,
with its corresponding rate containing an extra factor, namely the
Bose-Einstein factor.
Also in this case inelastic intensities are given by $\delta$-peaks.
Finally, the multiphonon contribution is obtained from the remaining
terms of the Taylor expansion or, equivalently, by subtraction of the
zero and single-phonon contributions.

Since we are interested in phonon lineshapes, a damping mechanism has
to be introduced in a theoretical model in order to replace the
$\delta$-peaks by Lorentzian lineshapes.
The goal in next Section is to assume a simple model, the so-called
Caldeira-Leggett (single-bath) model.
The corresponding Hamiltonian \cite{caldeira} has been widely used and
applied in the literature to a large variety of physical processes in
presence of dissipation \cite{weiss}.
More specifically, this approach has been extensively used in the
so-called Kramers' turnover problem \cite{pollak1a,pollak1b} describing
the escape from a metastable potential, in the vibrational dephasing
problem of small molecules in liquids \cite{pollak2}, and in diffusion
problems \cite{salva1a,salva1b,salva1c,salva2a,salva2b} with no
application to lineshapes.


\section{A simple model for phonon lineshapes:
The Caldeira-Leggett model}
\label{sec3}

Consider the damping of a single phonon (system) excited by some
external source and coupled to a thermal bath at a given temperature
(heat bath).
This dissipation mechanism is not very sensitive to the particular
loss process involved and, therefore, the heat bath can be assumed as
consisting of an infinite number of independent oscillators (phonon
field) in thermal equilibrium at the surface temperature.
Thermal fluctuations in the bath or reservoir, which feed with noise
the excited phonon, are related to the friction coefficient or damping
constant through the well-known fluctuation-dissipation theorem.
The system is usually represented by a harmonic oscillator weakly
coupled to the reservoir and undergoing an energy flow towards the
reservoir. Conversely, the reservoir fluctuations also couple back
into the system.
The driving force (He atoms) exciting the phonon is assumed to be
a linear perturbation and, at zero order, no influence in the
subsequent phonon dynamics is expected.
From the linear response theory, the dynamical susceptibility will be
that of a particle subject to a one-dimensional harmonic potential
\cite{salva2a,salva2b}.

Taking into account the comments above, the full Hamiltonian describing
the process can be written as \cite{caldeira,weiss}
\begin{equation}
\label{eq:H}
 H = H_S + H_R + H_{SR},
\end{equation}
where $H_S$ describes the Hamiltonian accounting for the system
dynamics (i.e.\ the excited phonon), $H_R$ is the Hamiltonian
associated with the thermal bath and $H_{SR}$ represents the
coupling between system and bath.
The so-called Caldeira-Leggett Hamiltonian has the structure given
by (\ref{eq:H}) and is usually expressed in the form
\begin{equation}
\label{eq:CLH}
 H = \frac{p^2}{2 m} + V(q)
  + \sum_{i=1}^N \left[ \frac{p_i^2}{2 m_i} +
   \frac{1}{2} \ \! m_i \left( \omega_i x_i
    - \frac{c_i}{m_i \omega_i} \ \! q \right)^2 \right] ,
\end{equation}
where $(q,p)$ and $m$ represent the system variables (coordinate and
conjugate momentum) and its mass, respectively; $V(q)$ is the external
potential acting on the uncoupled system, which here we assume to be
harmonic; $(x_i,p_i)$ stand for the variables of the $i$th bath
oscillator, characterized by a mass $m_i$ and a frequency $\omega_i$;
and $c_i$ is the coupling constant between the $i$th bath oscillator
and the system.

Within this single-bath model, tracing over the bath degrees of freedom
in the equations of motion (expressed in the Heisenberg picture) makes
evident the damped motion undergone by the system coordinate
\cite{weiss}.
This tracing gives rise to a quantum generalized Langevin equation,
\begin{equation}
\label{eq:GLE}
 m \ddot q + \frac{\partial V(q)}{\partial q} + m \int_0^t
  \gamma(t-t') \ \! \dot{q}(t') \ \! dt' = \xi (t) ,
\end{equation}
which describes the evolution of the system coordinate ($q$) under the
influence of the external harmonic potential $V(q)$, a friction kernel
$\gamma (t)$ and a random force $\xi (t)$ coming from the bath thermal
motion.
In this model, the friction kernel (phonon friction) is
\begin{equation}
 \label{eq:gt}
 \gamma (t) = \frac{1}{m}
  \sum_{i=1}^N \frac{c_i^2}{m_i \omega_i^2} \ \! \cos \omega_i t ,
\end{equation}
while the spectral density characterizing the bath oscillators is
\begin{equation}
\label{eq:J}
 \rho (\omega) = \pi \sum_{i=1}^N \frac{c_i^2}{2 m_i \omega_i}
  \delta (\omega - \omega_i) .
\end{equation}
On the other hand, the random force reads as
\begin{equation}
\label{eq:xit}
 \xi (t) = \sum_{j=1}^N  c_j
  \left\{ \left[ x_j (0) - \frac{c_j}{m_j \omega_j^2} \ \! q(0) \right]
   \cos \omega_j t
   + \frac{p_j(0)}{m_j \omega_j} \ \! \sin \omega_j t \right\} ,
\end{equation}
which satisfies the conditions for a Gaussian white noise provided the
friction is Ohmic, i.e.\ $\gamma (t) = 2 \gamma \delta (t)$.
In this case, Eq.~(\ref{eq:GLE}) becomes a standard quantum Langevin
equation,
\begin{equation}
\label{eq:LE}
 m \ddot q + \frac{\partial V(q)}{\partial q} + m
  \gamma \ \! \dot{q}(t) = \xi (t) .
\end{equation}

Quantum-mechanically, the Hamiltonian~(\ref{eq:CLH}) for a harmonic
potential $V(q) = m \omega_0^2 q^2 / 2$ can be expressed in terms of
system $(a, a^\dag)$ and bath $(b_j, b_j^\dag)$ operators as
\begin{eqnarray}
\label{eq:QCLH}
 \fl
 H & = & \hbar \omega_0 (a^{\dag}  a + 1/2) +  \sum_{j=1}^N
  \hbar \omega_j (b_j^{\dag} b_j + 1/2)
  + \hbar \sum_{j=1}^N \kappa_j
  (a  b_j^{\dag} + a^{\dag} b_j + a^{\dag} b_j^{\dag} + a b_j)
 \nonumber \\ \fl
 & &
  + \frac{\hbar}{4 \mu
      \omega_0} ({a^{\dag}}^2 + a^2 + 2 a^{\dag} a +1) \sum_{j=1}^N
        \frac{c_j^2} {m_j \omega_j^2} ,
\end{eqnarray}
with
\begin{equation}
 \kappa_j = \frac{c_j}{2 \sqrt{\mu m_j \omega_0 \omega_j}}
\end{equation}
accounting for the strength of the bilinear coupling between the system
and the bath oscillators.
As is apparent, two quantum transitions in the system are allowed
according to the existing second order terms.
The Hamiltonian (\ref{eq:QCLH}) can be diagonalized via a normal mode
transformation with mass weighted coordinates.
The corresponding normal-mode Hamiltonian can be then written as a
sum of $N$+1 oscillators \cite{pollak1a,pollak1b,pollak2},
\begin{equation}
\label{eq:NM}
 H = \frac{1}{2} \ \! P^2 + \frac{1}{2} \ \! w_0^2 Q^2
  + \sum_{j=1}^N \left( \frac{1}{2} \ \! P_j^2
   + \frac{1}{2} \ \! w_j^2 Q_j^2 \right) ,
\end{equation}
where $(w_0, w_j)$ denote the new system and bath frequencies
(eigenvalues) and $(Q,P)$ and $(Q_j,P_j)$ are the new system
and bath variables, respectively.
For Ohmic friction, the explicit expression of $w_0$ is
\begin{equation}
\label{eq:w0}
 w_0 = \pm \sqrt{\omega_0^2 - \gamma_0^2 /4} + i \gamma_0 / 2
  = \pm {\bar w}_0 + i \gamma_0 / 2 .
\end{equation}
Similarly, for $w_j$,
\begin{equation}
\label{eq:wj}
 w_j = \pm \sqrt{\omega_j^2 - \gamma_j^2 /4} + i \gamma_j / 2
  = \pm {\bar w}_j + i \gamma_j / 2 .
\end{equation}
The real parts in both cases give renormalized frequencies,
${\bar w}_0$ and ${\bar w}_j$, and the imaginary ones the damping
constant or phonon friction coefficient.
Notice that the new frequencies are no longer a simple linear function
of the friction coefficient, as it usually happens when only single
quantum transitions are allowed.
Furthermore, after diagonalization, we have a set of $N$+1 damping
phonons.
This procedure is equivalent to write Hamiltonian (\ref{eq:NM}) in
terms of real frequencies $\bar{w}_k$ [25,26].
Hence, in terms of operators, the Hamiltonian (\ref{eq:NM}) can be
recast as
\begin{equation}
\label{eq:NMa}
 H = \sum_{k=0}^N \hbar \bar{w}_k ({\bar a}_k^\dag {\bar a}_k + 1/2) ,
\end{equation}
where $({\bar a}^\dag, {\bar a})$ are related to the new phonon
variables $(Q_k,P_k)$.
The time evolution of such operators is given by
\begin{equation}
\label{eq:anm}
 {\bar a}_k (t) = {\bar a}_k (0) \ \! e^{i w_k t} ,
\end{equation}
with the minus sign usually chosen for the real part of the frequencies
in (\ref{eq:w0}) and (\ref{eq:wj}).

In order to obtain the phonon lineshapes, we proceed as previously
shown, although now the vibrational amplitude of the $j$th scattering
center is given in terms of the new normal modes or damping phonons,
expressed through $({\bar a}_k^\dag, {\bar a}_k)$ as
\begin{equation}
\label{eq:ujnm}
 {\bf u}_j (t) = \sum_{{\bf Q}, \nu}
  {\bar \lambda}_{{\bf Q},\nu}
  {\bf e}({\bf Q},\nu) \ \! e^{-\gamma_\nu ({\bf Q})t/2}
  \Big[ {\bar a}_{{\bf Q}, \nu} (t)
   e^{- i {\bf Q} \cdot {\bf R}_j}
 + {\bar a}_{-{\bf Q},\nu}^\dag (t)
   e^{i {\bf Q} \cdot {\bf R}_j} \Big] ,
\end{equation}
where, for simplicity, we keep the same notation $({\bf Q}, \nu)$ for
quantum numbers, as in (\ref{eq:uj}), and
\begin{equation}
 {\bar \lambda}_{{\bf Q},\nu} = \sqrt{ \frac{\hbar}{2 (N+1)
  {\bar w}_\nu ({\bf Q})}}  .
\end{equation}
Note that now the time-dependent operators are thus given by
(\ref{eq:anm}) and the corresponding frequencies by (\ref{eq:w0}) and
(\ref{eq:wj}), unlike Eq.~(\ref{eq:at}), where frequencies were real.
In this way, (\ref{eq:uplj}) acquires formally the same expression,
\begin{equation}
\label{eq:uplj1}
 \Upsilon_{lj} ({\bf k},t) =
    \frac{1}{2} \langle [{\bf k} \cdot {\bf u}_l(0)]^2 \rangle
  + \frac{1}{2} \langle [{\bf k} \cdot {\bf u}_j (t)]^2 \rangle
  - U_{jl} ({\bf k},t) ,
\end{equation}
but with the displacements being given by (\ref{eq:ujnm}).
This fact has important consequences: though the first term in
(\ref{eq:uplj1}) is $W$, as in (\ref{eq:DW}), the second one is no
longer independent of time (it depends on time as $e^{- \gamma_{\nu}
({\bf Q}) t}$) and, therefore, will not contribute as a new $W$ term.
Regarding the third term, it is also time-dependent, containing the new
extra factors $e^{- \gamma_\nu ({\bf Q}) t / 2}$, $e^{i {\bar w}_{\nu}
({\bf Q}) t}$ and the complex conjugate of the latter.
In particular, this displacement autocorrelation function, given above
by (\ref{eq:ujlT}), can now be expressed taking into account the new
normal modes as
\begin{eqnarray}
\label{eq:qjlTnm}
 U_{jl}({\bf k},t) & = & \sum_{{\bf Q}, \nu}
  {\bar \lambda}_{{\bf Q},\nu}^2 \ \!
   [{\bf k} \cdot {\bf e}({\bf Q},\nu)]^2
  \Big\{ n_{\nu}({\bf Q})
   e^{i {\bf Q} \cdot ({\bf R}_l - {\bf R}_j)} e^{-i {\bar w}_{\nu}
  ({\bf Q}) t} e^{- \gamma_{\nu} ({\bf Q}) t / 2}
 \nonumber \\ & &
  + [ n_{\nu} ({\bf Q}) + 1 ]
   e^{- i {\bf Q} \cdot ({\bf R}_l - {\bf R}_j)}
    e^{i {\bar w}_{\nu} ({\bf Q}) t}
     e^{- \gamma_{\nu} ({\bf Q}) t / 2} \Big\} ,
\end{eqnarray}
where ${\bar w}_{\nu} ({\bf Q})$ can be identified by ${\bar w}_k$.
Observe that this equation is similar to (\ref{eq:ujlT}) except for
the presence of damping factors.
Therefore, phonon lineshapes will be obtained from the time Fourier
transform, as given by (\ref{eq:skw}), but considering the damping
normal modes in terms of $({\bf Q}, \nu)$ and $w_{\nu} ({\bf Q})$.
The interest here relies on the fact that lineshapes can be obtained as
a function of the new system and bath frequencies issued from a given
diagonalization due to the presence of the heat bath.
This approach thus allows us to consider the excited phonon and the
phonon bath on {\it equal footing}.

Specifically, the phonon lineshapes obtained from the time Fourier
transform (\ref{eq:skw}) read as
\begin{eqnarray}
\label{eq:skw1}
 \fl S({\bf k},\omega) & = & \frac{1}{2 \pi \hbar} \ \! e^{-W({\bf k})}
  \sum_{j,l} \prod_{{\bf Q},\nu} \sum_{n,m,p=0}^{\infty}
  ( \frac{-1}{2})^p
  \frac{{\bar \lambda}_{{\bf Q},\nu}^{2(n+m+p)}}{n! m! p!} \ \!
  [{\bf k} \cdot {\bf e}({\bf Q},\nu)]^{2(n+m+p)}
   \nonumber \\ \fl & &
 \times  n_{\nu}({\bf Q})^n [n_{\nu} ({\bf Q}) + 1]^m
 [2 n_{\nu}({\bf Q}) +1]^p
 e^{i {\bf Q} \cdot ({\bf R}_l - {\bf R}_j) n}
  e^{- i {\bf Q} \cdot ({\bf R}_l - {\bf R}_j) m}
  \nonumber \\ \fl & &
 \times \frac{(n+m+2p) \gamma_{\nu}({\bf Q})}
  {[\omega-(m-n) {\bar w}_{\nu}({\bf Q})]^2
   + [(n+m+2p) \gamma_{\nu}({\bf Q})/2]^2} .
\end{eqnarray}
In principle, this expression contains the main ingredients of the
phonon relaxation dynamics.
For a given phonon, (\ref{eq:skw1}) consists of an infinite sum of
weighted Lorentzian functions which describes all phonon creation and
annihilation events.


\section{Discussion and applications}
\label{sec4}

Some consequences of (\ref{eq:skw1}) can be discussed at different
levels:

\vspace{.25cm}
\noindent
(a) {\it Damping phonons in clean surfaces: positions and lifetimes}
\vspace{.25cm}

The first point to be stressed from (\ref{eq:skw1}) is the frequency
position and lifetime of each (creation or annihilation) phonon event.
The frequency position is given by (\ref{eq:w0}) and involves the
phonon friction. Thus, the nominal value of the phonon frequency
$\omega_0$ or $\omega_j$ has to be replaced by a
renormalized frequency due to the coupling with the phonon bath.
The phonon lifetime is directly related to such a friction,
$\hbar / \gamma_{\nu}({\bf Q})$.
The different excitations of a certain phonon event are given by the
sums running over $n$ and $m$.
Each elementary contribution is expressed by a weighted Lorentzian
shape.
Obviously, the most prominent Lorentzian lineshape is related to the
most likely phonon event.
When $n = m$, we have the diffuse elastic peak.
This contribution arises from the creation and annihilation of the
same number of phonons and is essentially associated with single-phonon
exchanges (the multiphonon contribution is substantially smaller).
The product in the quantum numbers $({\bf Q},\nu)$ shows the
contributions of different excited phonons.
The multiphonon background is given by the different factors appearing
in (\ref{eq:skw1}).
The temperature dependence comes through the Bose-Einstein and DW
factors.
Regarding the symmetry of the total lineshape, and according to the
detailed balance condition, particle energy gains (which requires
the system to be in the higher energy state)
are less frequent than atom energy losses, in agreement with the
Boltzmann population factor.

\vspace{.5cm}
\noindent
(b) {\it Damping phonons and the DW factor}
\vspace{.25cm}

In (\ref{eq:skw1}), the DW factor appears as $e^{-W}$ instead of the
standard factor $e^{-2W}$.
This is a consequence of the diagonalization carried out in the
Caldeira-Leggett Hamiltonian (\ref{eq:CLH}).
The normal modes are now damping phonons given by (\ref{eq:anm}) and,
therefore, the square thermal averages of the displacements are no
longer independent of time.
As is well known, this factor is a measure of the effect of thermal
motions in reducing the periodicity of the lattice.
It is also a measure of the number of phonons involved in a scattering
event.
In this formalism, it is clearly seen that there is a new contribution
to the lifetimes (the sum running over $p$) independent of the creation
or annihilation of phonons.
This contribution comes from the whole damping lattice.

\vspace{.5cm}
\noindent
(c) {\it Adsorbates and phonons}
\vspace{.25cm}

The diffuse elastic intensity of the He atoms scattered at large angles
away from the specular direction provides very detailed information on
the mobility of adsorbates on surfaces.
At low coverages, the interaction among adsorbates can be ignored,
thus allowing to work within the so-called {\it single-adsorbate
approximation} describing the quasielastic and low frequency
vibrational lineshapes in terms of the {\it motional narrowing}
effect \cite{josele}.
The relaxation processes of creation or annihilation adsorbate events
are due to the coupling of the adsorbate modes (considered as the
system) to the phonon substrate (single-bath model).
Thus, (\ref{eq:skw1}) is also valid for describing the relaxation
dynamics of an excited single adsorbate with a renormalized frequency
given by (\ref{eq:w0}).
Furthermore, it can be considered as a generalization of previous
lineshapes obtained in a more phenomenological way \cite{josele}.

Based on the transition matrix formalism, Manson and Celli\cite{celli1}
proposed a quantum diffuse inelastic theory for small and intermediate
coverages of adsorbates on the surface by ignoring multiple scattering
effects of He atoms.
The dynamical structure factor is then obtained by assuming all the
lattice vibrational modes ($N_{ph}$) and point-like scattering centers
($N_{ad}$) satisfying the harmonic approximation with a given frequency
distribution function.
Therefore, following the same type of arguments, we could assume two
independent, uncorrelated baths to describe diffusion of interacting
adsorbates: the first bath consists of the surface phonons and the
second bath is formed by $N_{ad}$ adsorbates which obviously changes
with the surface coverage given by experimental conditions
\cite{ruth,salva2a,salva2b}.
The same procedure can now be used to obtain lineshapes of adsorbates;
in particular, the lowest frequency mode or frustrated traslational
mode.
To this end, we have to generalize the Hamiltonian (\ref{eq:CLH}) to
two baths, one for phonons and the other for adsorbates.
In this two-bath model (at a given coverage), we take one adsorbate as
the tagged particle or system, while the remaining ones constitute the
second bath described by $M$ harmonic oscillators.
Thus, when a given adsorbate is excited the coupling to phonons and
adsorbates simultaneously governs to the relaxation dynamics.
In this way, the corresponding total Hamiltonian in one dimension will
read as \cite{salva2a,salva2b}
\begin{eqnarray}
 H & = & \frac{p^2}{2m} + V(q) \nonumber \\
  & & + \sum_{i=1}^{N_{ph}}
 \left[ \frac{p_i^2}{2 m_i}+ \frac{m_i}{2}
  \left( \omega_i x_i - \frac{c_i}
    {m_i \omega_i} \ \! q \right)^2  \right]
   \nonumber \\
     & & + \sum_{j=1}^{N_{ad}}
 \left[ \frac{{\tilde p}_j^2}{2 {\tilde m}_j} + \frac{{\tilde m}_j}{2}
  \left( {\tilde \omega}_j {\tilde x}_j
  - \frac{d_j}{{\tilde m}_j {\tilde \omega}_j} \ \! q \right)^2
   \right] ,
\label{HCL2}
\end{eqnarray}
where now the tilde magnitudes refer to the second bath of $N_{ad}$
adsorbates, which are also taken as harmonic oscillators.
The $c_i$ and $d_i$ coefficients give the coupling strengths between
the adsorbate (system) and the substrate phonons or other adsorbates,
respectively.
The spectral density for the two baths is defined analogously to the
single-bath model,
\begin{equation}
  J(\omega) = \frac{\pi}{2}
 \sum_{i=1}^{N_{ph}} \frac{c_i^2}{m_i \omega_{i}^2}
  \ \! \delta (\omega - \omega_i)
   + \frac{\pi}{2}
 \sum_{j=1}^{N_{ad}} \frac{d_j^2}{{\tilde m}_j {\tilde \omega}_j^2}
  \ \! \delta (\omega - {\tilde \omega}_j) ,
 \label{SD2}
\end{equation}
but now it is split up into two terms, one spectral density associated
with the surface phonons and another one with the adsorbates.
In a similar way, the friction functions are defined as in
(\ref{eq:gt}), but with the spectral density being (\ref{SD2}).
The total friction function $\eta (t)$ also splits into two terms,
one due to the phonons, $\gamma (t)$, and another due to the collisions
with the the adsorbates or {\it collisional friction}, $\lambda (t)$:
$\eta (t) = \gamma (t) + \lambda (t)$ \cite{salva2a,salva2b}.
After diagonalization of Hamiltonian (\ref{HCL2}) and if Ohmic
friction is assumed, the new renormalized frequencies of the set of
$N_{ph} + N_{ad}$
oscillators are expressed in a similar way to Eqs. (\ref{eq:w0})
and (\ref{eq:wj}) but now in terms of $\eta$ instead of $\gamma$.
A straightforward generalization of lineshapes accounting for both
phonon and adsorbate (low-frequency modes) excitations at the same
time is then easily obtained. Again, this theoretical procedure is more
general than that previously reported based on phenomenological
arguments \cite{ruth}.

\newpage
\vspace{.5cm}
\noindent
(d) {\it The convolution problem. Some simple applications}
\vspace{.25cm}

As a final remark, we would like to mention that, taking into account
all the previous information, one could carry out the calculation of
the convolution of phonon and adsorbate lineshapes in order to extract
reliable lifetimes and damping coefficients or phonon frictions in a
more careful way.
Thus, as an illustration, now we will focus on experimental lineshapes
obtained from typical time-of-flight measurements from He atom-surface
scattering: an isolated frustrated translational ($T$) mode peak and
the masking of a $T$-mode by a substrate Rayleigh phonon.
In particular, we will consider He scattering off a Cu(001) surface
with a 3\% of CO coverage at low (surface) temperatures \cite{exp},
displaying the results in terms of the energy transferred ($\Delta E$)
in the scattering process.
Experimentally, the intrinsic peak width is of the same order of
magnitude as the apparatus energy resolution.
Therefore, quantitative reliable results can only be extracted by using
deconvolution techniques.
Thus, in the case of low surface temperatures, one proceeds
(numerically) by assigning a single Lorentzian function to the
intrinsic lineshape and convoluting it with the apparatus response
function, which is usually assumed to be a Gaussian function.
Then, by means of standard least squares fitting techniques, the
resulting curve is optimized until it fits the experimental one,
extracting finally from this procedure the good Lorentzian lineshape.
In the applications below, we have used this procedure rather than
Eq.~(\ref{eq:skw1}), since we are within a low temperature regime and
only one Lorentzian function is going to contribute.
Information about position of the maximum and full width at half
maximum is processed in terms of renormalized frequencies according
to (\ref{eq:w0}) and (\ref{eq:wj}).
As the surface temperature, more and more Lorentzian functions have
to be added in order to determine the lineshape, in agreement with
(\ref{eq:skw1}).

\begin{figure}
 \begin{center}
 \epsfxsize=8cm {\epsfbox{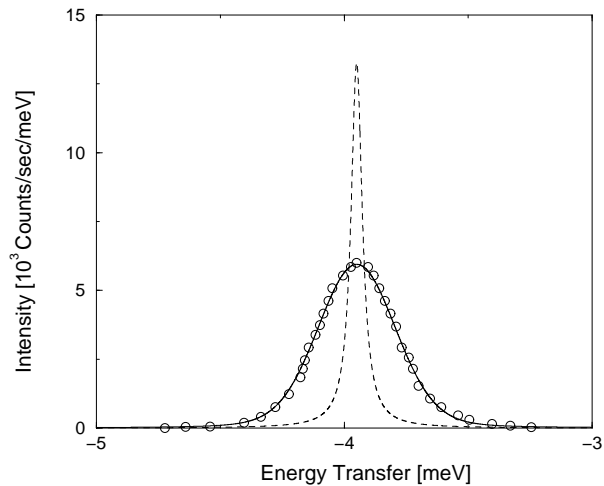}}
 \caption{\label{fig1}
  Comparison between experimental data ($\circ$) \cite{exp} and
  convolution results (solid line) for the creation of the $T$-mode
  peak ($\Delta E \approx -4$~meV) at $T_s = 60$~K.
  The dashed line represents the best-fitting Lorentzian function
  whose position and FWHM are determined in such a way that, once
  the Lorentzian is convoluted with the (Gaussian) apparatus response
  function, it best fits the experimental data.}
 \end{center}
\end{figure}

In Fig.~\ref{fig1} we show the result obtained (solid line) from the
convolution of a single Lorentzian function with the Gaussian apparatus
response function, i.e., a typical Voigt profile, which describes the
creation of a $T$-mode peak at $\Delta E = \hbar {\bar w}_0 \approx
-4$~meV and a surface temperature $T_s = 60$~K.
The dashed line represents the best-fitting Lorentzian function whose
position and full width at halh maximum (FWHM) are determined, as
mentioned above, by convoluting the Lorentzian with the apparatus
response function and then finding the best fit to the experimental
data (here, they have been extracted from Ref.~\cite{exp} and are
represented with open circles).
As can be seen, the agreement between the experimental data \cite{exp}
and our fitting is fairly good.
The FWHM is around 0.38~meV (damping constant) and $\omega_0$ can be
calculated from (\ref{eq:w0}).
As mentioned above, as the surface temperature increases, the number of
Lorentzian terms to be considered in (\ref{eq:skw1}) will also increase
in order to reproduce a wider and lower convolution profile, which can
be later compared with the experimental peak.

\begin{figure}
 \begin{center}
 \epsfxsize=8cm {\epsfbox{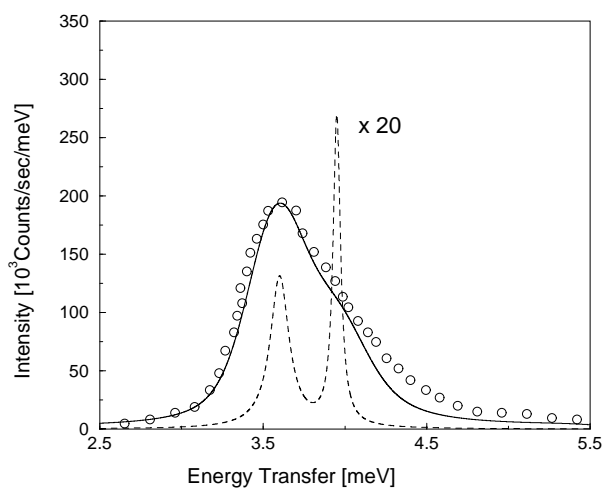}}
 \caption{\label{fig2}
  Comparison between experimental data ($\circ$) \cite{exp} and
  convolution results (solid line) for the overlapping of the peaks
  corresponding to the annihilation of both a substrate Rayleigh
  phonon (at $\Delta E \approx 3.5$~meV) and the $T$-mode of the
  adsorbate (at $\Delta E \approx 4$~meV) at $T_s = 60$~K \cite{exp}.
  The dashed lines represent the best-fitting Lorentzian functions
  whose positions and FWHMs are determined in such a way that, once
  the Lorentzians are convoluted with the (Gaussian) apparatus
  response function, they best fit the experimental data.}
 \end{center}
\end{figure}

In Fig.~\ref{fig2}, the results from the convolution (solid line)
describe the case where the peaks corresponding to the annihilation of
both a substrate Rayleigh phonon (at $\Delta E \approx 3.5$~meV) and
the $T$-mode of the adsorbate (at $\Delta E \approx 4$~meV) are very
close.
Here, again, the surface temperature is $T_s = 60$~K, which allows us
to consider only one Lorentzian function (dashed lines) for each event,
the excitation of the substrate phonon and the adsorbate $T$-mode.
As can be seen, we find a fairly good agreement between our
theoretical results and the experimental data (open circles)
\cite{exp}.
Both peaks are so close that they overlap, which leads the Rayleigh
phonon peak, more intense, to mask the peak corresponding to the
$T$-mode.
This is in a sharp contrast with the relative height of the
corresponding Lorentzians, which is lower for the former.
This overlapping of the two Lorentzian functions (dashed lines) thus
explains the formation of the shoulder observed at larger $\Delta E$.

\ack

We would like to thank Profs.\ V.\ Celli and J.\ R.\ Manson for very
interesting and stimulating comments and discussions about this work.
This is a very good opportunity to express our deep gratitude
to Vittorio and Dick to be always scientific references in the
atom-surface scattering field.
This work is dedicated to both of them.

This work has been supported in part by the Ministerio de Ciencia e
Innovaci\'on (Spain) under Project FIS2007-02461.
R.M.-C.\ thanks the Royal Society for a Newton Fellowship.
A.S.\ Sanz thanks the Consejo Superior de Investigaciones Cient\'{\i}ficas
for a JAE-Doc contract.


\end{document}